\documentclass[journal]{IEEEtran}

\ifCLASSINFOpdf
   \usepackage[pdftex]{graphicx}
\else
\fi

\usepackage[ruled,linesnumbered]{algorithm2e}
\usepackage{mathrsfs}
\usepackage{cite}
\usepackage{amsmath,amsthm,amsfonts}
\usepackage{amssymb}
\usepackage{multirow}
\usepackage{booktabs}
\usepackage{array}
\usepackage{makecell}
\usepackage{stfloats}
\usepackage{epstopdf}
\usepackage{epsfig}
\usepackage{subfigure}
\usepackage{graphicx}
\makeatletter
\thm@headfont{\sc}
\makeatother

\usepackage{url}

\usepackage{color}
\usepackage[noend]{algpseudocode}
\usepackage{ tipa }

\begin{document}

\title{Sampling Subgraph Network with Application to Graph Classification}

\author{Jinhuan~Wang,
        Pengtao~Chen,
        Bin~Ma,
        Jiajun~Zhou,
        Zhongyuan~Ruan,
        Guanrong~Chen,~\IEEEmembership{Fellow,~IEEE},
        and~Qi~Xuan,~\IEEEmembership{Member,~IEEE},

%
%
%
%
\IEEEcompsocitemizethanks{

\IEEEcompsocthanksitem J. Wang, P. Chen, J. Zhou, Z. Ruan, and Q. Xuan are with the Institute of Cyberspace Security, College of Information Engineering, Zhejiang University of Technology, Hangzhou 310023, China (e-mail: JinhuanWang@zjut.edu.cn;
Pengt.Chen@gmail.com; jjzhou@zjut.edu.cn; zyruan@zjut.edu.cn; xuanqi@zjut.edu.cn).
\IEEEcompsocthanksitem B. Ma is with the Department of Electrical and Computer Engineering, University of Southern California, Los Angeles, CA, the United States (e-mail: binma@usc.edu).
\IEEEcompsocthanksitem G. Chen is with the Department of Electronic Engineering, City University of Hong Kong, Hong Kong SAR, China (e-mail: eegchen@cityu.edu.hk).
\IEEEcompsocthanksitem Corresponding authors: Qi Xuan.}
}

\maketitle

\IEEEtitleabstractindextext{
\begin{abstract}

Graphs are naturally used to describe the structures of various real-world systems in biology, society, computer science etc., where subgraphs or motifs as basic blocks play an important role in function expression and information processing. However, existing research focuses on the basic statistics of certain motifs, largely ignoring the connection patterns among them. Recently, a subgraph network (SGN) model is proposed to study the potential structure among motifs, and it was found that the integration of SGN can enhance a series of graph classification methods. However, SGN model lacks diversity and is of quite high time complexity, making it difficult to widely apply in practice. In this paper, we introduce sampling strategies into SGN, and design a novel sampling subgraph network model, which is scale-controllable and of higher diversity. We also present a hierarchical feature fusion framework to integrate the structural features of diverse sampling SGNs, so as to improve the performance of graph classification. Extensive experiments demonstrate that, by comparing with the SGN model, our new model indeed has much lower time complexity (reduced by two orders of magnitude) and can better enhance a series of graph classification methods (doubling the performance enhancement). 
\end{abstract}

\begin{IEEEkeywords}
network sampling, subgraph network, feature fusion, graph classification, biological network, social network
\end{IEEEkeywords}}

\maketitle
\IEEEdisplaynontitleabstractindextext

\section{Introduction}

Networks or graphs are frequently used to capture various relationships that exist in the real world, and thus we witness the emergence of social networks~\cite{xuan2019self, kim2018social, 8281007}, traffic networks~\cite{ruan2019empirical, tang2020predictability, xu2020ge}, biological networks~\cite{walter2004visualization, wale2008comparison, zhou2020m}, literature citation networks~\cite{hosseini2018analysis, yasunaga2019scisummnet}, etc. The recently proposed graph representation methods allow us to better understand the structures of these networks and promote the development of various disciplines. Interestingly, the early graph embedding methods were benefited from natural language processing~\cite{MotifBasedAttention}, while now the graph neural networks (GNN) are used to successfully deal with visual semantic segmentation~\cite{2001.00335}. Furthermore, these graph embedding methods have made remarkable achievements in such areas as recommendation systems\cite{cheng2016wide,wang2019knowledge}, QA sites~\cite{zhang2018diffusion, fu2019nes}, and even drug discovery~\cite{jing2018deep, lane2018comparing}. In fact, network science, together with machine learning (especially deep learning), has made an important contribution to the development of cross-disciplines.

Subgraphs or motifs \cite{liu2019link,xuan2015temporal}, as basic building blocks, can be used to describe the mesoscale structure of a network. The networks constructed by different subgraphs may have vastly different topological properties and functions, and thus could be integrated into many graph algorithms to improve their performances.
For instance, after extracting the root subgraph with a modified skip-gram model, Narayanan et al.~\cite{1606.08928} proposed Subgraph2Vec as an unsupervised representation learning method, leading to good performance on graph classification. Ugander et al.~\cite{ugander2013subgraph} treated subgraph frequencies in social networks as local attributes and found that subgraph frequencies do provide unique insights for identifying social and graph structures of large networks. Inspired by neural document embedding models, Nguyen et al.~\cite{doi:10.1137/1.9781611975321.35} proposed the GE-FSG method, which adopts a series of frequent subgraphs as the inputs of the PV-DBOW model to obtain the entire-graph embeddings, achieving good performance in graph classification and clustering. 
These studies focus more on the basic statistics, e.g., the number of subgraphs, but lack analysis of the underlying structure among these subgraphs. The recently proposed subgraph network (SGN) model~\cite{xuan2019subgraph} takes the above issue into consideration and connects different subgraphs to construct a new network at a higher level. This process can be iterated to form a series of SGNs of different orders. It has been proven that SGNs can effectively expand the structural space and further improve the performance of network algorithms. 

However, SGN model has the following two shortages. First, the rule to establish SGN is deterministic, i.e., users can generate only one SGN of each order for a network. Such lack of diversity will limit the capacity of SGN to expand the latent structure space. Second, when the number of subgraphs exceeds the number of nodes in a network, the generated SGN can be even larger than the original network, which makes it extremely time-consuming to process SGNs of the higher-order, letting alone integrating these SGNs to design algorithms of better performances. On the other hand, it is noted that network sampling can increase the diversity by introducing the randomness, and meanwhile control the scale, providing an effective and inexpensive solution for network analysis. This merit thus is exactly complementary to the SGN model.

In this paper, we introduce network sampling into the SGN model, and proposes \underline{S}ampling \underline{S}ub\underline{G}raph \underline{N}etwork ($\text{S}^2$GN). In particular, we utilize the following four network sampling strategies, including random walk, biased walk, link selection, and spanning tree, to sample a subnetwork containing certain numbers of nodes and links, and then map the subnetwork to SGN based on certain rules. Network sampling and SGN construction can be used iteratively, so as to create a series of $\text{S}^2$GN of different orders, whose structural features can then be fused with those of the original network, so as to enhance a number of network algorithms. Specifically, we have the following contributions:
\begin{itemize}

\item We propose a new network model, sampling subgraph network ($\text{S}^2$GN), by introducing network sampling into SGN. Compared with SGN, our $\text{S}^2$GN can increase the diversity and decrease the complexity to a certain extent, benefiting the subsequent network algorithms.
\item We propose hierarchical fusion to fully utilize the structural information extracted from $\text{S}^2$GNs of different orders, generated by different sampling strategies, to enhance various graph classification algorithms based on manual attributes, Graph2Vec, DeepKernel, and CapsGNN.
\item We apply the new method to eight real-world network datasets, and our experimental results demonstrate the effectiveness and efficiency of $\text{S}^2$GN. The fusion of $\text{S}^2$GNs generated by different sampling strategies can increase the performance of graph classification algorithms in 30 out of 32 cases, with a relative improvement of 10.75\% on average (4.68\% for SGN). This value increases to 14.49\% (2.06\% for SGN) when only CapsGNN is considered, i.e., the combination of $\text{S}^2$GN-Fusion and CapsGNN achieves the $F_1$-$Score$ 80.98\% on average, greatly improving the graph classification performance. More remarkably, compared with SGN, generating S$^2$GNs needs much less time, reduced by almost two orders of magnitude.
\end{itemize}

The rest of the paper is structured as follows. In Sec.~\ref{sec:related}, we briefly describe the related work in network sampling and feature extraction. In Sec.~\ref{sec:SSGN}, we introduced the construction method of $\text{S}^2$GN. In Sec.~\ref{sec:experiments}, we give several feature extraction methods, which together with $\text{S}^2$GN are applied to eight real-world network datasets. Finally, we conclude the paper and highlight some promising directions for future work in Sec.~\ref{sec:Con}.

\section{Related work}\label{sec:related}

In this section, to supply some necessary background information, we give a brief overview of network sampling strategies and graph representation algorithms in graph mining and network science.

\subsection{Network Sampling}\label{sec:sampling}
Our work is closely related to the line of research in the network analysis based on sampling. Sampling methods in graph mining have two main tasks: generating node sequences and limiting the scale of the network. For the former, many studies utilize sampling strategies to extract node sequences to provide materials for subsequent network representation. Random walk~\cite{noh2004random} is one of the most famous node sampling methods, which has a wide influence in the field of graph mining~\cite{andersen2006local, fouss2007random}. For example, DeepWalk~\cite{perozzi2014deepwalk} combined the random walk with the language model in NLP, which was applied to node classification as a graph embedding method. In addition, Grover and Leskovec~\cite{grover2016node2vec} designed a biased walk mechanism based on random walk, which had a further improvement in node classification. Breadth-First Sampling~\cite{kurant2010bias} is a node sampling algorithm, which is biased to the nodes of high degrees and has been successfully applied in the measurement and topological analysis of OSNs. By limiting the scale of a network, Satuluri et al.~\cite{10.1145/1989323.1989399} sparsified graphs and achieved faster graph clustering without sacrificing quality. Moreover, sampling on graphs also has a wide spectrum of applications on network visualization~\cite{devi2019graph}. The sampling method can simplify the network while preserving significant structure information, which 
is of ultra importance in graph mining.


\subsection{Graph Representation}\label{sec:graphrep}

The most naive network representation method is to calculate graph attributes according to certain typical topological metrics~\cite{li2011graph}. Early graph embedding methods were considerably affected by NLP. For example, as graph-level embedding algorithms, Narayanan et al. proposed Subgraph2Vec~\cite{1606.08928} and Graph2Vec~\cite{narayanan2017graph2vec}, which achieve good performances on graph classification.

Another popular approach is to use graph kernel methods to capture the similarity between graphs.
Although representing networks well, they generally have relatively high computational complexity~\cite{li2011graph}. It is worth mentioning that the WL kernel~\cite{10.5555/1953048.2078187} was used to make the subgraph isomorphism check more effective. On this basis, Yanardag and Vishwanathan~\cite{yanardag2015deep} proposed an alternative kernel formulation termed as Deep Graph Kernel (DeepKernel) which achieved good performances on several datasets.

With the rise of spectral analysis of graph data in recent years, graph convolutional neural network (GCN) has been developed. It uses the Laplace decomposition of graphs to achieve convolutional operation in the spectral domain. Kipf et al.~\cite{kipf2016semi} used this neural network structure for semi-supervised learning, and achieved excellent results. Later, mathematical analysis on GCN went further and proved that the Laplacian decomposition used by GCN and Laplacian smoothing on images have mathematically equivalent forms~\cite{Li2018DeeperII}. At the same time, GCNs in the spatial domain have also been proposed. Inspired by the idea of convolution kernels in CNN, Mathias et al.~\cite{10.5555/3045390.3045603} proposed the method of PATCHY-SAN, which can determine the direction of the convolutions and the order of the nodes in the convolution window, and this model also achieved good results in graph classification. In this way, GCN treats the obtained information without weighting, i.e. the information of important neighbors and non-important neighbors will be put into the convolution layer in an unbiased manner. GAT overcomes this shortage by supplementing a self-attention coefficient before the convolution layer~\cite{veli2018graph}. Based on the newly proposed capsule network architecture, Zhang et al.~\cite{xinyi2018capsule} designed a CapsGNN to generate multiple embeddings for each graph, thereby capturing the classification-related information and the potential information with respect to the graph properties at the same time, which achieved the good performance.

Although the above graph representation methods have relatively high expressiveness and learning ability, largely improving the performance of graph classification, they do not have good interpretability, and in addition, they only rely on a single network structure, limiting their ability to exploit the latent structural space. Therefore, we generate multiple S$^2$GNs to fully expand the latent structural space, so as to enhance the network algorithms. Our experiments have demonstrated that S$^2$GNs can be naturally integrated with many graph representation methods by our feature fusion framework for the further improvement of their effectiveness.

\section{Methodology}\label{sec:SSGN}
We first briefly review SGN and the four network sampling methods. Then we introduce the framework to establish $\text{S}^2$GN. 

\subsection{Subgraph network}\label{sec:sgn}

\begin{figure}[h]
	\centering
	\includegraphics[width=0.9\linewidth]{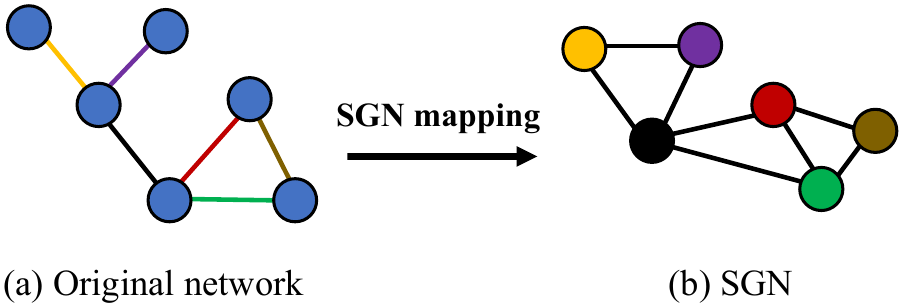}
	\caption{Schematic diagram of SGN construction.}
	\label{fig:SGN}
\end{figure}

Subgraph network (SGN)~\cite{xuan2019subgraph} is considered as a mapping function in network space. It provides a scalable model that transforms the original node-level network into a subgraph-level network. As shown in Fig.~\ref{fig:SGN}, the SGN in Fig.~\ref{fig:SGN} (b) can be obtained by SGN mapping from the original network in Fig.~\ref{fig:SGN} (a). One can see that the edges of different colors in (a) are mapped into the corresponding nodes in (b), which are naturally connected depending on whether they share the same node in the original network.

Formally, given an undirected network $G=(V,E)$ as an original network, where $V$ and $E$ are the node and edge sets, respectively. Let $V_{i}\subseteq V$ and $E_{i}\subseteq E$. Then, $g_{i}=(V_{i},E_{i})$ is a subgraph of $G$. The SGN, denoted by $G_{s} = \mathscr{L}(G)$, is a mapping from $G$ to $G_{s}=(V_{s},E_{s})$, where the node and edge sets are denoted by $V_{s}$=$\left\{g_{i}|i=0,1,2,...,n\right\}$ and $E_{s}\subseteq(V_{s}\times V_{s})$. If $g_{a}\cap g_{b} \neq \emptyset$, i.e., $g_{a}\cap g_{b} \in V$, in the original network, then they are connected in the SGN, i.e.,
$(g_{a}, g_{b})\in E_{s}$. It can be seen that the construction of SGN has three steps: (i) detect subgraphs $\left\{g_i \right\}$ from the original network; (ii) clear and define the connection rules between subgraphs; (iii) build SGN by leveraging the subgraphs.

For simplicity, here for the case of 1st-order SGN, denoted by SGN$^{\textbf{(1)}}$, pairwise linked nodes are chosen as building units, and the adjacent node pairs are connected. In this case, SGN$^{\textbf{(1)}}$ is equivalent to the line graph~\cite{harary1960some}, which reveals the topological interaction between edges of the original network. Fu et al.~\cite{8281007} used this method to map the original network to an SGN, and then used the node centrality in SGN to predict the weights of edges of the original network. As the SGN gradually maps to the higher-order network space, one can observe more abundant feature information. For example, the 2nd-order subgraph network, denoted by SGN$^{\textbf{(2)}}$, is obtained by repeating the mapping process on the SGN$^{\textbf{(1)}}$. The building unit of SGN$^{\textbf{(2)}}$ is a 2-hop structure (open triangle), which maintains the 2nd-order interactive information of the edge structures and can provide more insights about the local structure of a network~\cite{eckmann2002curvature}. To reduce the density of SGN, in the case of SGN$^{\textbf{(2)}}$, two building units are connected when they share the same edge. The latent structural information provided by higher-order SGNs may steadily diminish as the order increases. Therefore, SGN generally works best with the first two orders~\cite{xuan2019subgraph}.


\subsection{Network Sampling Strategies}\label{sec:ssd}

In this paper, we adopt the following four sampling strategies, including random walk, biased walk, link selection, and spanning tree, to design our $\text{S}^2$GN.

\textbf{Random walk.} Random walk ~\cite{pearson1905problem} can be used to obtain the co-occurrence relationship between nodes during network sampling.
A node in a network can be described by the wandering sequence starting from it. The wandering sequence obtained from the node contains both local and higher-order neighbors. When the wandering scope is extended to the graph level, one can peek into the topology of the whole network. 
In our model, given a network $G=(V, E)$, the random walk algorithm is described as follows:

\begin{itemize}
\item \emph{Start with an initial node $v^{0}\in V$.}
\item \emph{At step $i$, choose one neighbouring node $u$ $\in$ $\mathcal{N}(v^{i-1})$.}
\item \emph{Let $v^{i}$ $\leftarrow$ $u$ be the next node and get the edge $\widehat{E}$ $\leftarrow$ $\widehat{E}+\left\{(v^{i-1},v^{i})\right\}$.}
\item \emph{Repeat the steps until $|\widehat{E}|= |V|$.}
\end{itemize}
Node $v^{i}$ is generated by the following distribution:
\[P({v^{i}} = x|{v^{i - 1}} = m)=\left\{ \begin{array}{l}
{\textstyle{{{\alpha}} \over N}},if(m,x) \in E\\
0,otherwise
\end{array} \right.\]
where $\alpha$ is the transition probability between nodes $m$ and $x$, and $N$ is the normalizing constant. One can follow the above steps to simulate a random walk and get the final substructure $\widehat{G}=(\widehat{V},\widehat{E})$.

\textbf{Biased walk.} In the field of network science, biased walk ~\cite{azar1992biased} is different from the random walk where the probability of a potential new state is independent of external conditions. When the network is too complex to be analyzed by statistical methods, the biased walk provides an effective method for structural analysis by extracting the symmetry of an undirected network. The concept of the biased walk has attracted considerable attention, especially in the fields of transportation and social networks~\cite{adal2010biased}. Here, we adopt the walking mechanism of Node2Vec~\cite{grover2016node2vec}, where the homogeneity equivalence and structural equivalence of nodes are preserved by integrating the depth-first search and breadth-first search. Specifically, we adopt the 2nd-order random walk with parameters $p$ and $q$, which takes into account the topological distance between the next node and the previous node as well as the connectivity of the current node. Thus, the transition probability $\alpha$ between $v^{i}$ and $v^{i+1}$ is determined by
\[{\alpha _{(v^{i},v^{i+1})}} = {\omega _{pq}}(v^{i-1},v^{i+1}) = \left\{ \begin{array}{l}
\frac{1}{p},{d_{(v^{i-1},v^{i+1})}} = 0\\
1,{d_{(v^{i-1},v^{i+1})}} = 1\\
\frac{1}{q},{d_{(v^{i-1},v^{i+1})}} = 2
\end{array} \right.\]
where $v^{i-1}$, $v^{i}$, and $v^{i+1}$ are the previous, current, and next nodes, respectively, and $d_{(v^{i-1},v^{i+1})} \in (0,1,2)$ indicates the shortest path between $v^{i-1}$ and $v^{i+1}$. Note that $\alpha$ is equal to $\omega_{pq}$ when the network is unweighted. Various substructures of network can be obtained by controlling $p$ and $q$.


\begin{figure}[t]
    \centering
	\includegraphics[width=1.0\linewidth]{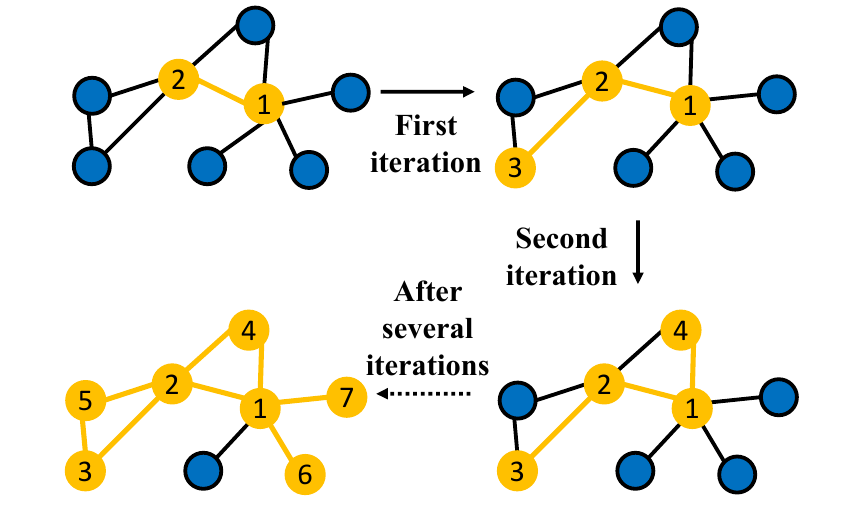}
	\caption{Illustration of the walk procedure in link selection.}
	\label{fig:LSpic}
\end{figure}

\textbf{Link selection.} We also propose a new edge-based sampling method, namely link selection. Given a network $G=(V, E)$, we first sample an initial edge $e^0=(v^{0}, v^{1})$, and then randomly select a node of this edge as the source node of the next sampling edge. The nodes of all the sampled edges form the source node pool $V_{pool}$ for the next sampling. The sampling process will not terminate until the stop condition is met. The substructures after this sampling strategy are obtained by a diffuse search from a central edge, which ensures the acquisition of important network structures to a certain extent. As shown in Fig.~\ref{fig:LSpic}, the node pair (1,2) is selected as the initial edge and then we can get the substructure that contains nodes (1,2,3) after one iteration through node "2" and get an expanding substructure that contains nodes (1,2,3,4) after second iteration through another node "1". After several iterations, one can get the final substructure, which contains 7 nodes and 8 edges while the program satisfies the stop condition.

\begin{itemize}
\item \emph{Start with an initial edge $e^{0}=(v^{0}, v^{1})\in E$, and let $V_{pool}=\left\{v^{0}, v^{1}\right\}$, $E_{pool}=\left\{e^{0}\right\}$.}
\item \emph{At step $i$, choose one node $u\in V_{pool}$. }
\item \emph{Let $u^{i}$ $\leftarrow$ $u$ be the next start node and select an edge $(u^{i},u^{i+1}) \notin E_{pool}$.}
\item \emph{Update $V_{pool}$ $\leftarrow$ $V_{pool}+\left\{u^{i+1}\right\}$ and get the edge pool $E_{pool}$ $\leftarrow$ $E_{pool}+\left\{(u^{i},u^{i+1})\right\}$.}
\item \emph{Repeat the above steps until $|E_{pool}|= |V|$.}
\end{itemize}

Note that $(u^{i},u^{i+1})$ has the same transition probability with the random walk, and $V_{pool}$ and $E_{pool}$ are the node and edge sets of the final substructure $\widehat{G}$. This method differs from random walk in that it can search the network on the basis of the current substructure rather than a single node, which can reduce the appearance of a chain structure to a greater extent.

\begin{figure} [htpb]
    \centering
	\includegraphics[width=1.0\linewidth]{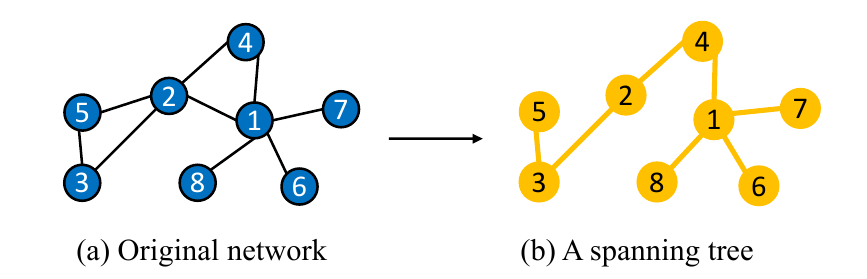}
	\caption{The substructure obtained by spanning tree.}
	\label{fig:STpic}
\end{figure}

\textbf{Spanning tree.} A spanning tree~\cite{dey2019new} is a minimally connected substructure that contains all nodes in the graph, as shown in Fig.~\ref{fig:STpic}. Different spanning trees can be obtained by traversing from different nodes. Here we randomly select a node as the initial node. The maximum and minimum spanning trees are unified without considering the edge weights. In this section, we use the typical Kruskal algorithm~\cite{najman2013playing} to generate spanning trees and the weight values of edges are all set to 1.

\subsection{Framework for Constructing $\text{S}^2$GN}\label{sec:ssgn}

Most real-world networks have large scale and complex structure. Typically, SGN could be even larger and denser, making the follow-up network algorithms less efficient. It may also introduce extra noisy structural information, disturbing the network algorithms to a certain extent. In view of this, we focus on optimizing the SGN model and propose a framework for constructing a sampling subgraph network (S$^2$GN) by integrating different network sampling methods.
The pseudocodes of constructing S$^2$GN and sampling substructures are given in Algorithms~\ref{alg:1} and \ref{alg:2}, respectively. In Algorithms~\ref{alg:1}, GetMaxSubstructure(·) is to obtain the maximally connected substructure of original network if it is not connected; NodeRanking(·) is to rank the input nodes; SGNAlgorithms(·) is to construct SGNs. GetNextEdgeWithStrategy(·) in Algorithms~\ref{alg:2} is to get the next edge according to a given sampling strategy.


\begin{figure*}[!t]
	\centering
	\includegraphics[width=1\linewidth]{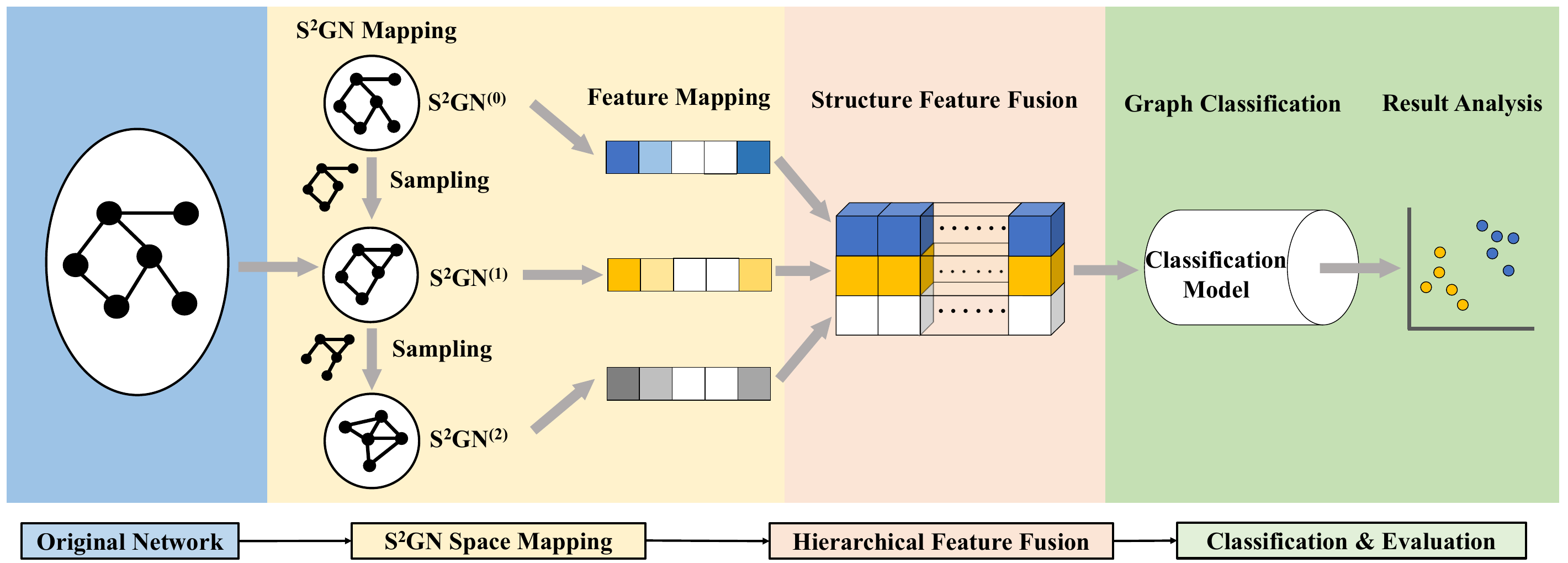}
	\caption{The overall framework of the S$^2$GN algorithm for network structure feature fusion.}
	\label{fig:SSGN}
\end{figure*}

\begin{algorithm}[!t]
\caption{\textbf{Construction of S$^2$GN.}}
\LinesNumbered
\label{alg:1}
\KwIn{A network $G$($V$,$E$) with node set $V$ and link set $E\subseteq(V\times{V})$\;
Sampling strategy $f_s(\cdot)$\;
The order of SGN $h$.}
\KwOut{S$^2$GN, denoted by $G_s$($V_s$,$E_s$).}
Initialize a temporary object $G_s$ = $G$\;
\While {$h$}
{ \If {the $G_s$ is not full-connected}
    {
    GetMaxSubstructure($G_s$)\;
    Initial node $u$ = NodeRanking($V_s$)\;
    Get sampling substructure $\widehat{G_s}$ through executing  Algorithm \ref{alg:2}\; 
    $G_{sgn}$ = SGNAlgorithms($\widehat{G_s}$)\;
    $G_s$ $\leftarrow$ Relabeled($G_{sgn}$)\;
    }
  \Else {Repeat 5-8\;}
  $h = h - 1$\;
}
\Return $G_s$($V_s$,$E_s$)
\end{algorithm}


\begin{algorithm}[!t]
\caption{\textbf{Sampling substructure.}}
\LinesNumbered
\label{alg:2}
\KwIn{A network $G$($V$,$E$)\;
Source node $u$\;
Sampling walks $l$.}
\KwOut{Sampling substructure, denoted by $\widehat{G_s}$=$g$($\widehat{v}$,$\widehat{e}$).}
Let $v_0$=$u$, initial $walk_v$ to [$v_0$], $walk_e$ to $\emptyset$\;
Select first edge $e_1$ with a given probability of sampling strategy\;
Append the $v_1$ = $dst$($e_1$) to $walk_v$, $e_1$ to $walk_e$\;
\For {$i=2$ to $l-1$}
{   $cur_v$ = $walk_v$[-1], $cur_e$ = $walk_e$[-1]\;
    $e_i$ = GetNextEdgeWithStrategy($cur_v$, $cur_e$)\;
    Append $e_i$ to $walk_e$, $v_i$=$dst(e_i)$ to $walk_v$\;}
    $\widehat{v}$ = $walk_v$, $\widehat{e}$ = $walk_e$\;
\Return $\widehat{G_s}$=$g$($\widehat{v}$,$\widehat{e}$)
\end{algorithm}

In general, S$^2$GN can be constructed in three steps: source node selection, sampling substructure and S$^2$GN construction, which are introduced in the following.

\begin{itemize}
\item \textbf{\textbf{Source node selection}}: There are many ways to choose the initial node: (i) Randomly select a node as the source node; (ii) Select an initial node according to its importance measured by closeness centrality~\cite{okamoto2008ranking}, K-shell~\cite{lu2016vital}, PageRank~\cite{langville2004deeper} or others. In this paper, we use the K-shell method in order to capture the key structure more likely.
\item \textbf{\textbf{Sampling substructure}}: After the initial source node is determined, a substructure can be obtained by conducting a certain sampling strategy to extract the main context of the current network. According to different sampling strategies, diverse sampling substructures can be generated, reflecting the different aspects of the original network and further benefiting the subsequent network algorithms.
\item \textbf{\textbf{S$^2$GN construction}}: Based on the sampling substructure, we use SGN model to construct S$^2$GN. Note that network sampling and SGN are adopted iteratively so as to get the S$^2$GNs of higher orders. This method can control the size of S$^2$GNs and meanwhile increase their diversity. Therefore, compared with SGN, the S$^2$GN could further enhance both efficiency and effectiveness of the subsequent network algorithms.
\end{itemize}

\begin{figure*}[!t]
	\centering
	\includegraphics[width=1\linewidth]{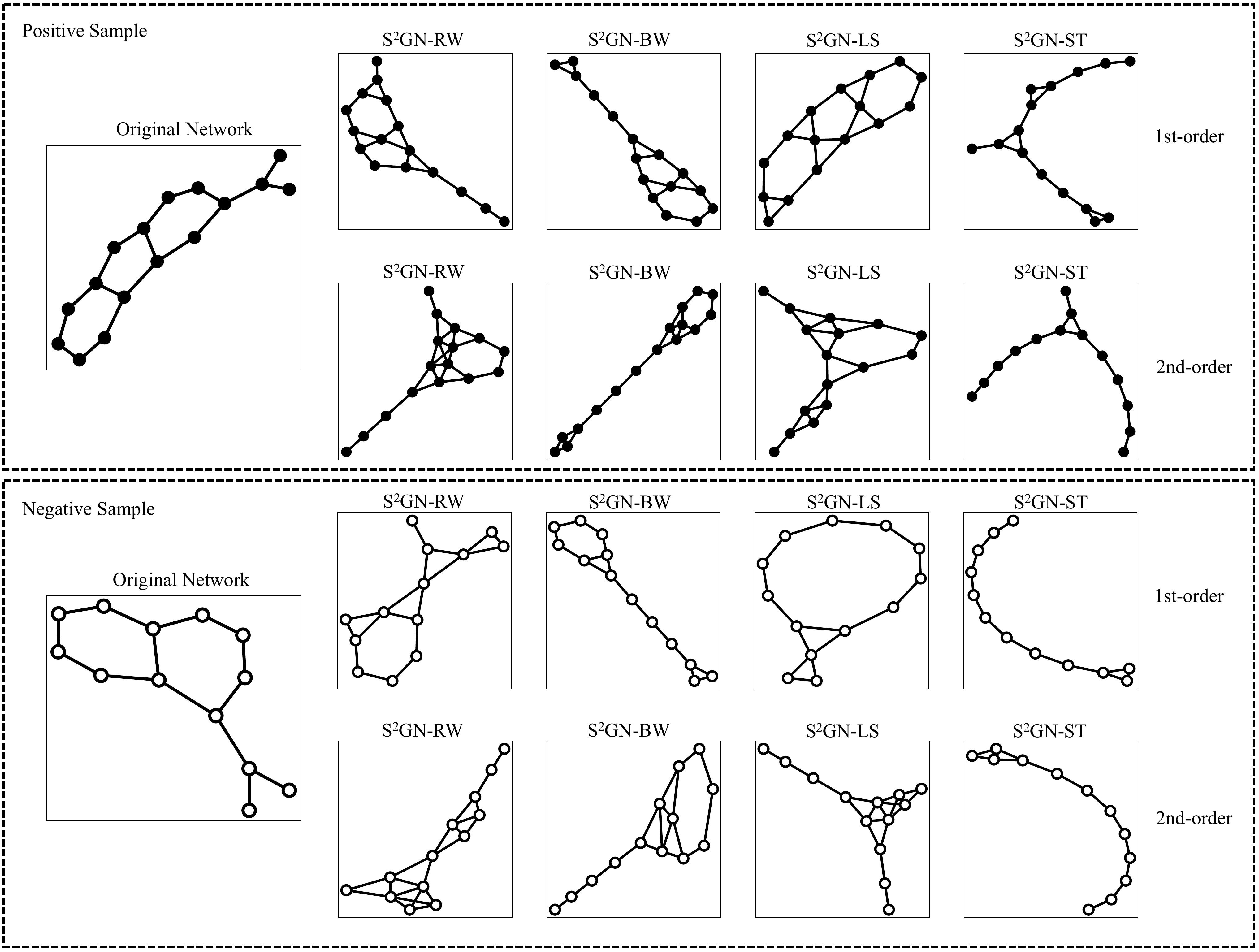}
	\caption{Visualization of 1st-order and 2nd-order S$^2$GNs using four network sampling strategies on positive and negative samples from the MUTAG dataset.}
	\label{fig:NetV}
\end{figure*}

    

Now, we use various feature extraction methods to get structural features from S$^2$GNs of different orders, which are first fused and then used to establish the graph classification models. The overall framework of S$^2$GN construction for structural feature space expansion is shown in Fig.~\ref{fig:SSGN}. Note that, generally, information fusion tries to integrate information from multiple aspects to improve algorithm performance, which has a wide range of applications in practice. For instance, in speech recognition, the visual features of the lip motion are fused with the speech signal features to predict the words expressed~\cite{zhou2019modality}. In image recognition, Xuan et al.~\cite{8219720} developed a multistream convolutional neural network to automatically merge the features of multi-view pearl images, so as to improve the accuracy of pearl classification. In this paper, we use different sampling strategies to capture the structural features from different aspects. As an example, we visualize different 1st-order and 2nd-order S$^2$GNs generated by the four network sampling strategies on positive and negative samples from the MUTAG dataset, as shown in Fig.~\ref{fig:NetV}. It can be seen that the S$^2$GNs generated by different sampling strategies have quite different structures, and the structural difference between the positive and negative samples may be enlarged in S$^2$GNs. Therefore, it can be expected that the fusion of these diverse S$^2$GNs could improve the performance of graph classification.


\section{Experiments}\label{sec:experiments}

Now, we compare S$^2$GN and SGN models on their abilities to enhance graph classification based on four feature extraction methods. We first introduce the datasets, followed by the feature extraction methods and the parameter setting. After that, we show the experimental results with discussion.

\subsection{Datasets}

We test our S$^2$GN method on eight real-world network datasets, as introduced in the following. IMDB-BINARY is about social networks,  while the others are about bio- and chemo-informatics networks. The basic statistics of these datasets are presented in Table~\ref{data}.

\begin{table}[!h]\renewcommand{\arraystretch}{1.2}
	\newcommand{\tabincell}[2]{\begin{tabular}{@{}#1@{}}#2\end{tabular}}
	\caption{Basic statistics of eight datasets. $N_G$ is the number of graphs, \#$C_{max}$ is the number of graphs belonging to the largest class, $N_C$ is the number of classes, and \#Nodes and \#Edges are the average numbers of nodes and edges, respectively, of the graphs in the dataset.}
    \centering
	\begin{centering}
		\begin{tabular}{l|ccccc}
			\hline\hline
			Dataset&  $N_G$ & \#$C_{max}$ & $N_C$  & \#Nodes & \#Edges  \tabularnewline
			\hline
			MUTAG& \tabincell{c}{188}  &125  &2  &18  &20\tabularnewline
			PTC& \tabincell{c}{344}   &192    &2  &14  &14\tabularnewline
			PROTEINS& \tabincell{c}{1113} &663   &2  &39  &73\tabularnewline
			ENZYMES& \tabincell{c}{600}  & 100  &6  &32  &63\tabularnewline
			NCI1& \tabincell{c}{4110}  &2057    &2  &30  &32\tabularnewline
			NCI109& \tabincell{c}{4127} &2079   &2  &30  &32\tabularnewline
            IMDB-BINARY& \tabincell{c}{1000} &500   &2  &20  &193\tabularnewline
            D$\&$D & \tabincell{c}{1178} &691   &2  &284  &716\tabularnewline
			\hline\hline		
		\end{tabular}		
	\end{centering}	
	\label{data}
\end{table}

\begin{itemize}
\item \textsc{\textit{MUTAG}}~\cite{debnath1991structure} contains 188 mutagenic aromatic and heteroaromatic compounds, with nodes and edges representing atoms and the chemical bonds between them, respectively. They are labeled according to whether there is a mutagenic effect on a special bacteria.

\item \textsc{\textit{PTC}}~\cite{toivonen2003statistical} includes 344 chemical compound graphs, with nodes and edges representing atoms and the chemical bonds between them, respectively. Their labels are determined by their carcinogenicity for rats. 

\item \textsc{\textit{PROTEINS}}~\cite{borgwardt2005protein} comprises of 1113 graphs. The nodes are Secondary Structure Elements (SSEs) and the edges are neighbors in the amino-acid sequence or in the 3D space. These graphs represent either enzyme or non-enzyme proteins. 

\item \textsc{\textit{ENZYMES}}~\cite{nr} contains 600 protein tertiary structures, and each enzyme belongs to one of the 6 EC top-level classes. 

\item \textsc{\textit{NCI1}} \& \textsc{\textit{NCI109}}~\cite{wale2008comparison} comprise of 4110 and 4127 graphs, respectively. The nodes and edges represent atoms and chemical bonds between them, respectively. They are two balanced subsets of the datasets of chemical compounds screened for the activities against non-small cell lung cancer and ovarian cancer cell lines, respectively. The positive and negative samples are distinguished according to whether they are effective against cancer cells. 

\item \textsc{\textit{IMDB-BINARY}}~\cite{nguyen2018learning} is about movie collaboration including 1000 graphs, which is collected from IMDB and contains lots of information about different movies. Each graph is an ego-network, where nodes represent actors or actresses and edges indicate whether they appear in the same movie. Each graph is categorized into one of the two genres (Action and Romance).

\item \textsc{\textit{D$\&$D}}~\cite{dobson2003distinguishing} contains 1178 graphs of protein structures. A node represents an amino acid and edges are constructed if the distance between two nodes is less than 6 $\mathring{A}$. A label denotes whether a protein is an enzyme or non-enzyme.
\end{itemize}

\subsection{Feature Extraction Methods}\label{sec:fem}

We adopt four typical methods to generate graph representation, namely manual attributes, Graph2Vec, DeepKernel, and CapsGNN, which are introduced in the following.

\begin{itemize}
\item \textit{Attributes}: Here, we use the same 11 manual attributes as those introduced in \cite{xuan2019subgraph}, including the number of nodes, the number of edges, average degree, network density, average clustering coefficient, the percentage of leaf nodes, the largest eigenvalue of the adjacency matrix, average betweenness centrality, average closeness centrality, and average eigenvector centrality.

\item \textit{Graph2Vec}~\cite{narayanan2017graph2vec}: This is the first unsupervised embedding approach for an entire network, which is based on the extending word-and-document embedding techniques that has shown great advantages in natural language processing (NLP).

\item \textit{DeepKernel}~\cite{yanardag2015deep}: This method provides a unified framework that leverages the dependency information of sub-structures by learning latent representations. The sub-structure similarity matrix, $\mathcal{M}$, is calculated by the matrix $\mathcal{V}$ with each column representing a sub-structure vector. Denote by $\mathcal{P}$ the matrix with each column representing a sub-structure frequency vector. According to the definition of kernel: $\mathcal{K} = \mathcal{P}\mathcal{M}\mathcal{P}^\mathrm{T} = \mathcal{P}\mathcal{V}\mathcal{V}^\mathrm{T}\mathcal{P}^\mathrm{T}=\mathcal{H}\mathcal{H}^\mathrm{T}$, one can use the columns in the matrix $\mathcal{H}=\mathcal{P}\mathcal{V}$ as the inputs to the classifier.

\item \textit{CapsGNN}~\cite{xinyi2018capsule}: This method was inspired by CapsNet~\cite{sabour2018matrix}, which adopts the concept of capsules to overcome the weakness of existing GNN-based graph embedding algorithms. In particular, CapsGNN extracts node features in the form of capsules and utilizes the routing mechanism to capture important information at the graph level. The model generates multiple embeddings for each graph so as to capture graph properties from different aspects.
\end{itemize}

\subsection{Parameter Setting}\label{sec:ps}
For source node selection, we choose the node of the largest K-shell~\cite{lu2016vital} as the source node for random walk (RW) and biased walk (BW), and choose the edge of the largest betweenness centrality as the source edge for link selection (LS). We randomly pick up a node as the source node for the spanning tree (ST) to increase the diversity of S$^2$GN, since the sampled subnetworks will be quite similar if we fix the source node for this method. Moreover, we set the two parameters of BW as $p=4$ and $q=1$.


\begin{table*}[!t]
\caption{Classification results measured by $F1$-$Score$ on eight datasets by using different feature extraction methods.}
\centering
\renewcommand\arraystretch{1.2}
\setlength{\tabcolsep}{0.7mm}{
\begin{tabular}{c|cccccccc|c}
\hline\hline
   \textbf{Algorithm} & \multicolumn{ 8}{c}{\textbf{Classification results} ($F1$-$Score$, \%)} \\
\hline
   \textbf{Attributes} & MUTAG & PTC & PROTEINS & ENZYMES & NCI1 & NCI109 & IMDB-BINARY & D$\&$D & Avg. \\
\hline
    Original & $86.58\pm{3.61}$  &$63.52\pm{4.55}$  &$78.30\pm{2.49}$  &$43.37\pm{2.29}$  & $67.48\pm{0.87}$  & $67.34\pm{1.25}$ &$73.00\pm{3.68}$ & $75.85\pm{1.61}$ &  69.43\\
\hline
       SGN & $91.58\pm{4.21}$ & $67.94\pm{6.36}$ & $79.46\pm{2.96}$ & $50.22\pm{2.91}$ & $69.84\pm{1.59}$ & $69.73\pm{1.97}$ & ${\bf77.65\pm{4.50}}$ & $76.65\pm{1.59}$  &  72.88\\
$RIMP$-SGN &5.78\%   &6.96\%   &1.48\%    &15.79\%   &3.50\%   &3.55\%   &6.37\%  &1.05\% &4.97\%\\
\hline
  S$^2$GN-RW & $90.53\pm{2.11}$ & $66.71\pm{1.62}$ & $77.76\pm{1.52}$ & $52.17\pm{2.36}$& $74.82\pm{0.69}$ & $73.96\pm{0.84}$ & $71.85\pm{2.74}$ & $77.37\pm{2.91}$ &  73.15 \\

  S$^2$GN-BW & $93.94\pm{2.37}$ & $69.11\pm{2.94}$ & ${\bf79.83\pm{1.46}}$ & $53.37\pm{2.78}$& $75.47\pm{0.98}$ & $73.99\pm{1.04}$ & $76.05\pm{1.29}$ & ${\bf77.75\pm{1.68}}$ &  74.94 \\

  S$^2$GN-LS & $89.21\pm{2.48}$ & $66.18\pm{2.55}$ & $78.57\pm{1.44}$ &$49.50\pm{2.14}$& $75.85\pm{1.03}$ & $74.65\pm{0.62}$ & $71.80\pm{2.53}$ & $76.91\pm{1.94}$ &  72.83 \\

  S$^2$GN-ST & $90.79\pm{2.12}$ & $70.44\pm{2.75}$ & $76.28\pm{1.89}$ & $45.33\pm{1.29}$& $72.25\pm{1.08}$ & $73.26\pm{0.76}$ & {\bf $77.60\pm{1.50}$} & $76.90\pm{2.47}$ &  72.81 \\
\hline
S$^2$GN-Fusion & ${\bf94.74\pm{1.84}}$ &  ${\bf72.06\pm{3.29}}$ & $79.14\pm{0.84}$ & ${\bf55.25\pm{1.90}}$& ${\bf76.03\pm{1.32}}$ & ${\bf74.89\pm{1.18}}$ & $76.97\pm{1.21}$ & $77.03\pm{2.46}$ &  ${\bf75.76}$ \\
$RIMP$-Fusion &9.42\%   &13.44\%   &1.07\%    &  27.39\%  &  12.67\%   & 11.21\%   & 5.44\%  & 1.56\%  & 9.12\%     \\
\hline\hline
   \textbf{Graph2Vec} & MUTAG &  PTC & PROTEINS & ENZYMES & NCI1 & NCI109 & IMDB-BINARY & D$\&$D & Avg. \\
\hline
Original &$83.15\pm{9.25}$  &$60.17\pm{6.86}$  &$73.30\pm{2.05}$  &$45.17\pm{2.73}$ &$73.22\pm{1.81}$ & $74.26\pm{1.47}$ & $62.47\pm{3.99}$ & $70.25\pm{2.18}$ & 67.75\\
\hline
       SGN & $86.84\pm{5.70}$ & $63.24\pm{6.70}$ & $74.44\pm{3.09}$ & $48.73\pm{2.56}$ & $76.64\pm{3.21}$ & $74.86\pm{2.76}$ & $70.65\pm{5.55}$ & $80.42\pm{3.06}$ & 70.73 \\
$RIMP$-SGN &4.44\%   &  5.10\%   &  1.56\%    &  7.88\%  &  4.67\%   &  0.81\%   &  13.09\%   &  14.48\%  & 4.39\%     \\
\hline
  S$^2$GN-RW & $80.26\pm{2.69}$ & $61.47\pm{2.06}$ & $76.37\pm{1.12}$ & $48.67\pm{2.53}$& $76.88\pm{1.35}$ & $74.39\pm{1.40}$ & $68.35\pm{1.57}$ & $81.86\pm{1.80}$ & 71.03\\

  S$^2$GN-BW & ${\bf86.84\pm{3.07}}$ & ${\bf64.71\pm{2.85}}$ & ${\bf77.13\pm{1.09}}$ & $52.33\pm{2.30}$ & $77.39\pm{1.12}$ & $75.69\pm{1.46}$ & $71.64\pm{2.00}$ & $82.12\pm{2.22}$ &  $73.48$\\

S$^2$GN-LS & $81.05\pm{2.57}$ & $62.35\pm{2.88}$ & $76.91\pm{2.21}$ &  $47.68\pm{1.73}$ & ${\bf79.18\pm{1.71}}$ & ${\bf77.42\pm{1.13}}$ & $67.25\pm{2.16}$ & $81.77\pm{1.98}$ & 71.70\\

 S$^2$GN-ST & $81.84\pm{2.99}$ & $63.97\pm{2.39}$ & $75.20\pm{2.15}$ & $49.87\pm{2.91}$ & $76.30\pm{1.21}$ & $72.95\pm{0.89}$ & $72.49\pm{2.11}$ & $74.92\pm{2.89}$ & 70.94 \\
\hline
S$^2$GN-Fusion & $81.73\pm{3.37}$ & $64.38\pm{2.42}$ & $75.10\pm{0.89}$ & ${\bf54.78\pm{2.29}}$ & $76.91\pm{0.72}$ & $75.72\pm{1.31}$ & ${\bf76.43\pm{2.17}}$ & ${\bf82.75\pm{2.79}}$ & ${\bf73.48}$\\
$RIMP$-Fusion &-1.71\%   &  7.00\%   & 2.46\%    &  21.28\%  &  5.04\%   &  1.97\%   &  22.35\%  &  17.79\%  &  8.46\% \\
\hline\hline
   \textbf{DeepKernel} & MUTAG & PTC & PROTEINS & ENZYMES & NCI1 & NCI109 & IMDB-BINARY & D$\&$D & Avg.  \\
\hline
Original &$82.95\pm{2.68}$  &$59.04\pm{1.09}$ &$73.30\pm{0.82}$  &$45.04\pm{3.73}$  &$67.06\pm{1.91}$   &$67.04\pm{1.36}$   &$67.50\pm{2.45}$ & $75.97\pm{1.91}$ & 67.24\\
\hline
       SGN & $93.68\pm{5.15}$ & $65.88\pm{5.05}$ & $76.78\pm{2.41}$  & $45.93\pm{3.75}$ & $70.26\pm{1.24}$ & $71.06\pm{1.61}$ & $75.70\pm{1.55}$ & $77.84\pm{2.08}$ & 72.14\\
$RIMP$-SGN &12.94\%   &  11.59\%   &  4.75\%    &  1.98\%  &  4.77\%   &  6.00\%   &  12.15\%  &  2.46\%  &  7.29\%\\
\hline
 S$^2$GN-RW & $93.68\pm{5.66}$ & $61.76\pm{3.77}$ & $75.80\pm{4.21}$ & $43.32\pm{3.64}$ & $69.15\pm{1.63}$ & $69.06\pm{1.70}$ & $72.30\pm{2.68}$ & $83.47\pm{1.00}$ & 71.07\\

 S$^2$GN-BW & $94.00\pm{5.43}$ &  $67.35\pm{4.48}$ &  $76.69\pm{2.97}$ & $47.75\pm{2.68}$ & $71.51\pm{1.38}$ & $69.83\pm{2.05}$ & $74.10\pm{3.33}$ & $81.57\pm{1.11}$ & 72.85\\

 S$^2$GN-LS & $93.68\pm{4.59}$ & $66.18\pm{4.21}$ & $76.16\pm{1.92}$ & $50.28\pm{3.04}$ & ${\bf71.55\pm{1.15}}$ & $70.19\pm{2.26}$ & $75.80\pm{3.43}$ & $\bf{83.98\pm{1.77}}$ & 73.48\\

 S$^2$GN-ST & $88.95\pm{3.68}$ & $65.29\pm{4.59}$ & $74.73\pm{4.54}$  & $48.02\pm{3.52}$ & $70.77\pm{1.20}$ & ${\bf71.04\pm{1.03}}$ & $75.90\pm{2.07}$ & $78.94\pm{1.38}$ & 71.71\\
\hline
S$^2$GN-Fusion & ${\bf94.73\pm{4.07}}$ & ${\bf70.88\pm{4.25}}$ & ${\bf77.14\pm{2.97}}$ & ${\bf52.21\pm{2.24}}$ & $71.06\pm{1.01}$ & $70.48\pm{1.22}$ & ${\bf76.50\pm{3.75}}$ & $83.77\pm{1.87}$ &  ${\bf74.60}$\\
$RIMP$-Fusion &14.20\%   &  20.05\%   &  5.24\%    &  15.92\%  &  5.96\%   &  5.13\%   &  13.33\%  &  10.27\%  &  10.94\%\\
\hline\hline
   \textbf{CapsGNN} & MUTAG &  PTC & PROTEINS & ENZYMES &  NCI1 & NCI109 & IMDB-BINARY & D$\&$D & Avg.  \\
\hline
Original &$86.32\pm{7.52}$&  $62.06\pm{4.25}$&  $75.89\pm{3.51}$  &$49.78\pm{3.02}$  & $78.30\pm{1.80}$&  $72.99\pm{2.15}$&  $72.71\pm{4.36}$ & $67.75\pm{2.57}$ & 70.73\\
\hline
       SGN & $89.47\pm{7.44}$ & $64.12\pm{3.67}$ & $76.34\pm{4.13}$  & $50.04\pm{2.70}$ & $78.61\pm{1.87}$ & $73.72\pm{2.39}$ & $76.47\pm{5.74}$ & $68.71\pm{1.91}$ & 72.19\\
$RIMP$-SGN &3.65\%   &  3.32\%   & 0.59\%    &  0.52\%  &  0.40\%   &  1.00\%   &  5.17\%   &  1.42\%  &  2.06\%\\
\hline
 S$^2$GN-RW & $88.70\pm{4.59}$ & $77.81\pm{4.96}$ & $84.73\pm{2.09}$ & $51.33\pm{1.14}$ & $74.23\pm{1.40}$ & $75.16\pm{1.39}$ & $92.50\pm{3.15}$ & $78.05\pm{1.91}$ & 77.81\\

 S$^2$GN-BW &  $92.63\pm{4.82}$ & $81.91\pm{5.45}$ & $84.10\pm{3.72}$ & $52.77\pm{2.11}$ & ${\bf78.83\pm{2.35}}$ & $75.25\pm{1.69}$ & $93.35\pm{1.12}$ & ${\bf78.66\pm{2.32}}$ & 79.69\\

 S$^2$GN-LS & $90.53\pm{4.59}$ & $79.11\pm{4.16}$ & $84.28\pm{1.96}$  & $52.17\pm{1.23}$ & $76.57\pm{1.26}$ &  $75.43\pm{1.46}$ & $93.92\pm{1.75}$ & $77.03\pm{1.52}$ & 78.63\\

 S$^2$GN-ST & $89.21\pm{5.73}$ & $78.67\pm{5.06}$ & $84.03\pm{2.58}$  & $52.56\pm{1.18}$ & $76.52\pm{1.42}$ & $75.16\pm{1.60}$ & $94.20\pm{1.26}$ & $72.31\pm{2.73}$ & 77.83 \\
\hline
       S$^2$GN-Fusion & ${\bf93.15\pm{4.11}}$ & ${\bf84.12\pm{6.47}}$ & ${\bf85.18\pm{1.84}}$ & ${\bf56.08\pm{3.15}}$ & $78.13\pm{2.27}$ & ${\bf78.23\pm{1.05}}$ & ${\bf95.10\pm{2.30}}$ & $77.85\pm{1.95}$ &  ${\bf80.98}$\\
$RIMP$-Fusion &7.91\%   &  35.55\%   &  12.24\%    &  12.66\%  &  -0.22\%   &  7.18\%   &  30.79\%  &  14.90\%  &  14.49\% \\
\hline\hline
\end{tabular}}
\label{Results}
\end{table*}

\begin{figure*}[!t]
	\centering
	\includegraphics[width=1\linewidth]{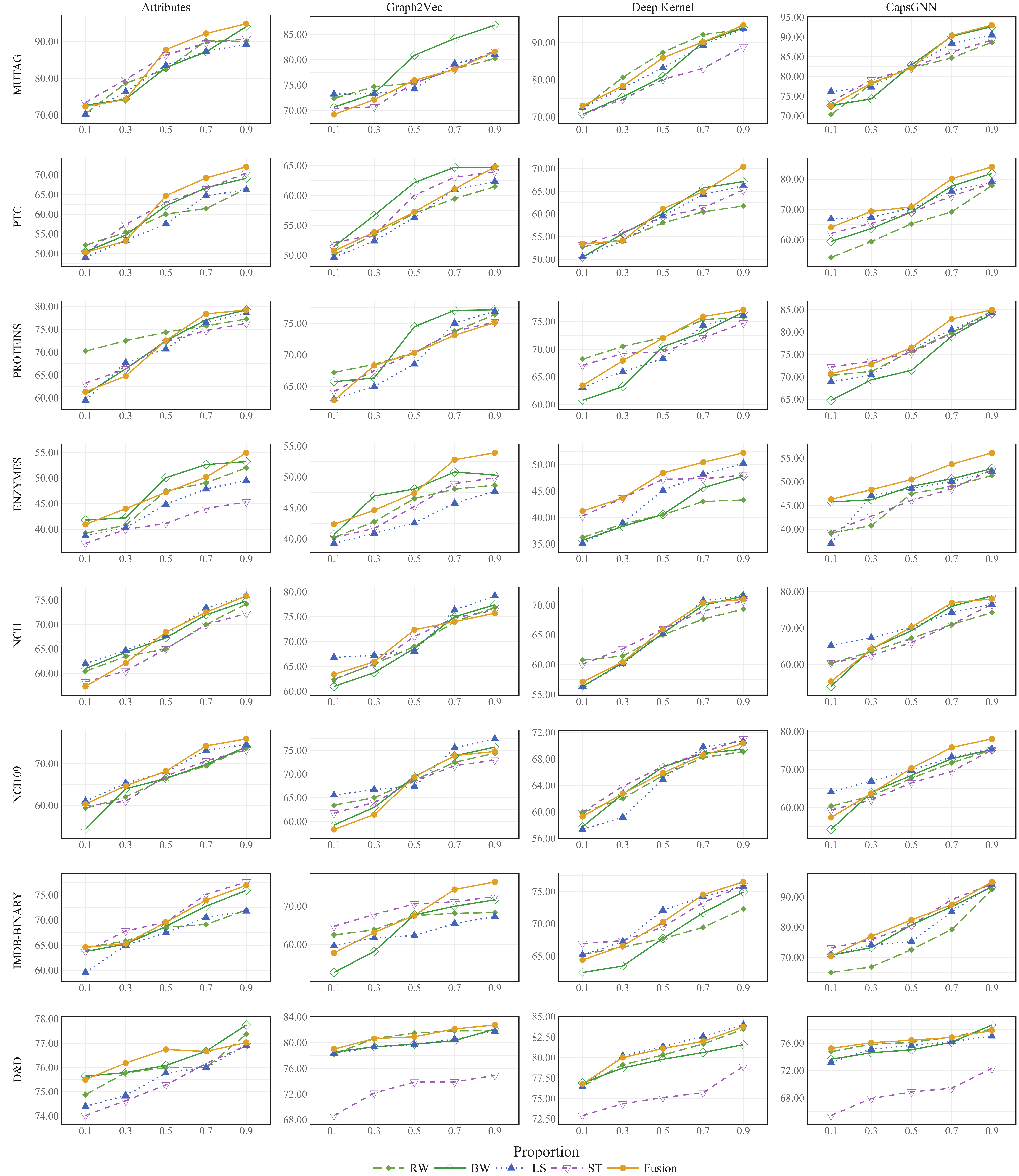}
	\caption{Average $F1$-$Score$ as functions of the training set size (represented by the fraction of samples in the training set), for various feature extraction methods on different datasets, based on RW, BW, LS, ST and Fusion, respectively.}
	\label{fig:plot}
\end{figure*}

In this study, for \emph{Graph2Vec}, the embedding dimension is adopted according to~\cite{narayanan2017graph2vec}. Since the embedding dimension is predominant for learning performances, a commonly-used value of 1024 is adopted. The other parameters are set to default values: the learning rate is set to 0.5, the batch size is set to 512 and the number of epochs is set to 1000. For \emph{DeepKernel}, according to ~\cite{yanardag2015deep}, the Weisfelier-Lehman subtree kernel is used to build the corpus and its height is set to 2. Furthermore, the embedding dimension is set to 10, the window size is set to 5 and skip-gram is used for the word2vec model. We adopt the default parameters for \emph{CapsGNN} and flatten the multiple embeddings of each graph as the input.

Without loss of generality, the well-known Random Forest is chosen as the classification model. Meanwhile, for each feature extraction method, the feature space is first expanded by using S$^2$GNs, and then the dimension of the feature vectors is reduced to the same value as that of the feature vector obtained from the original network using PCA in the experiments, for a fair comparison. Each dataset is randomly split into 8 folds for training and 2 fold for testing. Here, the $F1$-$Score$ is adopted as the metric to evaluate the classification performance:
\begin{equation}
F_{1}= \frac{2PR}{P + R}\,,
\end{equation}
where $P$ and $R$ are the precision and recall, respectively. In order to diminish the random effect of the fold assignment to some extent, the experiment is repeated 100 times and then the average $F_1$-$Score$ and its standard deviation are reported.

We further define the relative improvement rate (RIMP) of SGN or S$^2$GN model as
\begin{equation}
RIMP= (F1_{model}-F1_{ori})/F1_{ori}\,
\end{equation}
where $F1_{ori}$ and $F1_{model}$ refer to the $F1$-$Score$ of the graph classification algorithm without and with the SGN model (or S$^2$GN-Fusion model), respectively.

%
%
%
%
%

\subsection{Experimental Results}\label{sec:er}
We use the four network sampling strategies to generate sampling substructures, and further construct the corresponding 1st-order and 2nd-order S$^2$GNs, denoted by S$^2$GN-RW, S$^2$GN-BW, S$^2$GN-LS, and S$^2$GN-ST, respectively\footnote{It has been proven that the graph classification models can be significantly enhanced by appropriately using the structural information of the SGNs in the first two orders, while such gain will be reduced soon as more SGNs of higher orders are integrated~\cite{xuan2019subgraph}. This is why we only use the S$^2$GNs of the first two orders here.}. After that, we adopt the four feature extraction methods, namely manual attributes, Graph2Vec, DeepKernel, and CapsGNN, to get structural feature vectors. For each feature extraction method, we fuse the vectors generated from the different S$^2$GNs to a single vector. Finally, this vector is fed into the Random Forest model to produce the classification result. Note that we also produce the results for a single sampling strategy for a more comprehensive comparison. Here, a ten-fold cross-validation method is used to calculate $F1$-$Score$ of graph classification. To enrich the sampling structures and reduce the probability of sampling repetition, 10 sampling averaging processes were carried out for each sampling strategy.


\subsubsection{Enhancement on classification performance} 
The experimental results are shown in Table~\ref{Results}, where one can see that the four S$^2$GN models based on a single sampling strategy, i.e., S$^2$GN-RW, S$^2$GN-BW, S$^2$GN-LS, and S$^2$GN-ST, are comparable with the SGN model, which all produce similar classification results under different datasets and feature extraction methods. Interestingly, S$^2$GN-BW outperforms SGN in enhancing the classification models based on the four feature extraction methods in most cases, leading to a relative improvement of 4.52\% on average. Such results are consistent with the experience that Node2Vec is a powerful method to capture the structural properties of a network. Moreover, since different S$^2$GNs generated by different sampling strategies can capture the different aspects of a network, as visualized in Fig.~\ref{fig:NetV}, one may expect that the fusion of these S$^2$GNs can produce even better classification results. Indeed, we find that the fusion of $\text{S}^2$GNs increases the performance of the original graph classification algorithms in 30 out of 32 cases, with a relative improvement of 10.75\% on average (much better than 4.68\% by SGN). The value increases to 14.49\% (much better than 2.06\% by SGN) when only CapsGNN is considered. This result is quite impressive, since CapsGNN, together with $\text{S}^2$GN, achieves the state-of-the-art performance on PROTEINS and IMDB-BINARY datasets.

To address the robustness of our S$^2$GN model against the size variation of the training set, the $F_1$-$Score$ is calculated by using various sizes of training sets (from 10 to 90 percent, within a 20 percent interval). For each size, the training and test sets are randomly divided, which is repeated 100 times with the average result recorded. The results are shown in Fig.~\ref{fig:plot} for various feature extraction methods on eight datasets. It can be seen that still the curves of S$^2$GN-Fusion are relatively higher than those of S$^2$GNs generated by a single sampling strategy in most cases, indicating that the superiority of S$^2$GN-Fusion is robust in enhancing graph classification algorithms. In particular, such superiority seems much more significant when enhancing CapsGNN, which is interesting and may indicate that the potential of S$^2$GN-Fusion could be exploited further by connecting a better embedding method or end-to-end graph neural network, and meanwhile there could be much room for further improvement for graph classification. 


%



\subsubsection{Reduction of time complexity}
Note that one important motivation to introduce sampling strategies into SGN is to control the network size so as to improve the efficiency of the network algorithms based upon them. Therefore, here to address the computational complexity of our method, we record the average computational time of SGN and S$^2$GN generated by the four sampling strategies on the eight datasets, namely MUTAG, PTC, PROTEINS, ENZYMES, NCI1, NCI109, IMDB-BINARY, and D$\&$D. The results are presented in Table~\ref{Complexity}, where one can see that, overall, the computational time of S$^2$GN is much less than that of SGN for each sampling strategy on each dataset, decreasing from hundreds of seconds to less than 19 seconds. In fact, the computational time of S$^2$GNs generated by different sampling strategies is comparable to each other. Considering that S$^2$GN-Fusion method needs to generate all the four S$^2$GNs, its computational time is close to the sum of individual ones, which is still less than 25 seconds. Such results suggest that, by comparing with SGN, our S$^2$GN model can indeed largely increase the efficiency of the network algorithms.

In fact, we can estimate the time complexity of our model in theory. For random walk, it is a computationally efficient sampling method, which only requires $\mathcal{O}(|E|)$ space complexity to store the neighbors of each node in the graph. As for the time complexity, by imposing graph connectivity in the sample generation process, random walk provides a convenient mechanism to increase the effective sampling rate by reusing samples across different source nodes. For biased walk, we adopt the 2nd random walk mechanism of Node2Vec, where each step of random walk is based on the transition probability $\alpha$ which can be precomputed, so the time consuming of each step using alias sampling is $\mathcal{O}(1)$. Link selection broadens the scope of the start node at each step in the random walk process, thereby accelerating the time to reach the stop condition. Kruskal algorithm to generate spanning trees is a greedy algorithm, which has $\mathcal{O}(|E|log(|E|))$ time complexity. We know that the computational complexity of SGN$^{(1)}$ is $\mathcal{O}(|E|^2)$ and that of constructing SGN$^{(2)}$ is $\mathcal{O}(|E|^4)$. Our S$^2$GN model constrains the expansion of the network scale and reduces the cost of constructing SGNs to the fixed $\mathcal{O}(|E|^2)$. Thus, the time computational complexity $\mathcal{T}$ of our S$^2$GN model is $\mathcal{O}(|E|+|E|^2)\leq \mathcal{T}\leq \mathcal{O}(|E|log|E|+|E|^2|)$ according to the different sampling strategies, which is much lower than that of SGN.

\begin{table}[!t]
\caption{Average computational time to establish SGN and S$^2$GNs by the four sampling strategies on the eight datasets.}
\centering
\renewcommand\arraystretch{1.1}
\begin{tabular}{lccccc}
\hline\hline
\multirow{2}{*}{Time (Seconds)} & \multirow{2}{*}{SGN} & \multicolumn{4}{c}{S$^2$GN}    \\ \cline{3-6}
                            &                      & RW    & BW    & LS    & ST    \\
\hline
MUTAG                       & $1.58\times{10^2}$   & 0.677 & 0.252 & 0.600   & 0.090  \\
PTC                         & $1.93\times{10^3}$   & 1.216 & 0.804 & 1.170  & 0.607 \\
PROTEINS                    & $3.20\times{10^3}$   & 1.192 & 1.161 & 2.018 & 1.625 \\
ENZYMES                     & $3.97\times{10^3}$   & 1.284 & 1.230 & 2.106 & 1.598 \\
NCI1                        & $1.75\times{10^2}$   & 2.670 & 2.099 & 2.484 & 1.746 \\
NCI109                      & $1.75\times{10^2}$   & 2.682 & 2.114 & 2.495 & 1.749 \\
IMDB-BINARY                 & $1.11\times{10^4}$   & 1.478 & 1.580  & 1.256 & 1.106\\ 
D$\&$D                          & $7.90\times{10^2}$   & 2.701 & 3.162  & 18.32 & 0.805\\\hline
\end{tabular}
\label{Complexity}
\end{table}

\subsubsection{Visualization}

\begin{figure}[!t]
	\centering
	\includegraphics[width=1\linewidth]{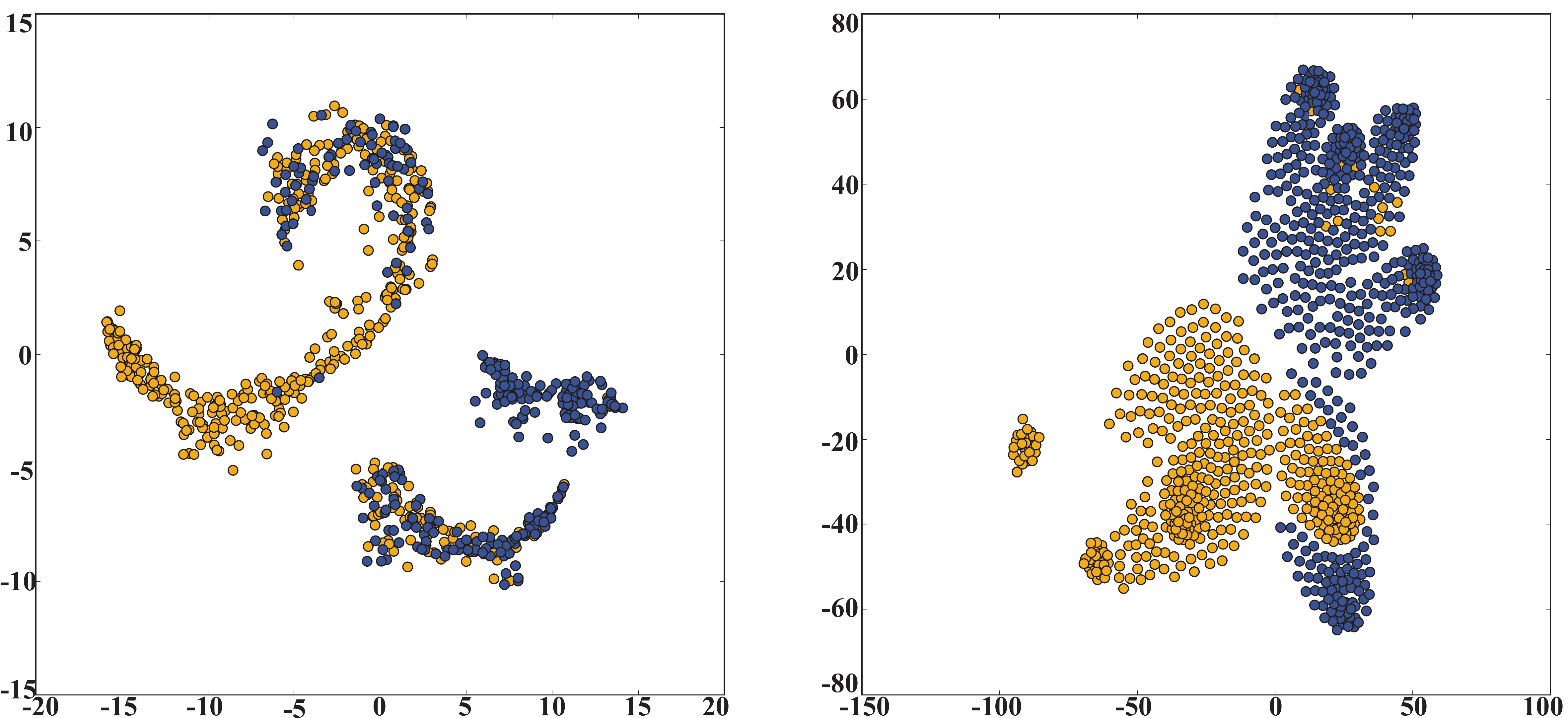}
	\caption{The t-SNE visualization of structural features using CapsGNN without (left) and with (right) S$^2$GN-ST. The same color of points represent the same class of graphs in IMDB-BINARY dataset.}
	\label{fig:TSNE}
\end{figure}


As a simple case study, we visualize the results of classification on IMDB-BINARY dataset based on CapsGNN method to verify the effectiveness of our S$^2$GN model. Here, we choose S$^2$GN-ST to visualize since this is the best S$^2$GN generated by the single sampling strategy that enhances the classification performance of CapsGNN most. As shown in Fig.~\ref{fig:TSNE}, the structural features are located in different places by utilizing t-SNE. The left shows the original classification result using CapsGNN without S$^2$GN-ST, while the right depicts the optimized distribution of the same dataset using CapsGNN with S$^2$GN-ST. One can see that the graphs in IMDB-BINARY dataset can indeed be distinguished by the original features of CapsGNN, but it appears that the distinction of graphs could become more explicit after hierarchical representation through network sampling and SGN mapping, demonstrating the effectiveness of our S$^2$GN model. 




\section{Conclusions}\label{sec:Con}

In this paper, we present a novel sampling subgraph network (S$^2$GN) model as well as a hierarchical feature fusion framework for graph classification by introducing network sampling strategies into the SGN model. Compared with the latter, the S$^2$GNs are of higher diversity and controllable scale, and thus benefit the network feature extraction methods to capture more various aspects of the network structure with higher efficiency. 

We use different sampling strategies, namely random walk (RW), biased walk (BW), link selection (LS), and spanning tree (ST), to generate the corresponding sampling subgraph networks S$^2$GN-RW, S$^2$GN-BW, S$^2$GN-LS, and S$^2$GN-ST, respectively. The experimental results show that, compared with SGN, S$^2$GN has much lower time complexity, which was reduced by almost two orders of magnitude, and meanwhile they have comparable effects on graph classification. In fact, the network algorithms based on S$^2$GN-BW behave even better than those based on SGN, although each sampling subnetwork is only a part of the original network.  More interestingly, when the features of all the four S$^2$GNs are fused and then fed into graph classification models, the classification performance can be significantly enhanced. In particular, when CapsGNN is used to extract the features of  these S$^2$GNs, we can achieve the-state-of-the-art results on the PROTEINS and IMDB-BINARY datasets.  


In the future, we will try more sampling strategies and then integrate them with SGN to generate more diverse S$^2$GNs; we will also apply our framework to more tasks beyond graph classification, such as link prediction, node classification, etc.

\section{Acknowledgments}\label{sec:ack}
The authors would like to thank all the members in the IVSN Research Group, Zhejiang University of Technology for the valuable discussions about the ideas and technical details presented in this paper. This work was partially supported by the National Natural Science Foundation of China under Grant 61973273, by the Zhejiang Provincial Natural Science Foundation of China under Grant LR19F030001, and by the Hong Kong Research Grants Council under the GRF Grant CityU11200317. 



\ifCLASSOPTIONcaptionsoff
  \newpage
\fi

\bibliographystyle{IEEEtran}
\bibliography{sample}

\end{document}